\documentclass[traditabstract]{aa} 
\usepackage{graphicx}
\usepackage{natbib}
\usepackage{txfonts}
\begin{document}
\title{The double degenerate system NLTT~11748\thanks{Based on observations at 
the European Organisation for Astronomical Research in the Southern Hemisphere,
Chile under programme ID 84.D-0862.}}
\author{A. Kawka\inst{1}
\and
S. Vennes\inst{1,2}
\and
T.R. Vaccaro\inst{2,3}
}

\institute{Astronomick\'y \'ustav, Akademie v\v{e}d \v{C}esk\'e republiky, Fri\v{c}ova 298, CZ-251 65 Ond\v{r}ejov, Czech Republic
\email{kawka,vennes@sunstel.asu.cas.cz}
\and
Visiting Astronomer, Kitt Peak National
Observatory, National Optical Astronomy Observatory, which is operated by the
Association of Universities for Research in Astronomy (AURA) under cooperative
agreement with the National Science Foundation.
\and
Department of Physics \& Astronomy, Francis Marion University, Box 100547, 
Florence, SC 29501, USA
\email{tvaccaro@fmarion.edu}
}

\date{Received ; accepted }

\abstract{We show that the extremely low-mass white dwarf NLTT~11748 ($0.17\,M_\odot$) is in a close binary
with a fainter companion. We obtained a series of radial velocity measurements of the low-mass white dwarf using
the H$\alpha$ core and determined an orbital period of $5.64$ hours. The
velocity semi-amplitude ($K=274.8$ km s$^{-1}$) and orbital period imply that it is a degenerate star, and that the minimum
mass for the companion is $0.75\ M_\odot$ (assuming a mass of $0.167\,M_\odot$ for the primary).
Our analysis of Balmer line profiles shows
that a $0.75\ M_\odot$ white dwarf companion does not contribute more than
2\% or 5\% of the flux (V-band) for helium- or hydrogen-rich surfaces, respectively. 
The kinematics of the system suggest that it belongs to the Galactic halo.
}


\maketitle

\section{Introduction}

The high proper-motion star NLTT~11748 was identified as an extremely low mass (ELM) white dwarf by 
\citet{kaw2009}. 
The first ELM white dwarfs were
discovered to be companions to neutron stars \citep[e.g.,][]{van1996,bas2003},
but recent follow-up observations of ELM white dwarfs identified in    
colourimetric or proper motion surveys (e.g., SDSS, NLTT) often find them
to be in a close binary with a white dwarf companion
\citep[e.g., SDSS~J125733.63$+$542850.5,][]{bad2009,kul2010,mar2010}.
The companions of several other ELM white dwarfs found to be in close binary systems 
are yet to be formally identified, but they are likely to be more massive
white dwarfs \citep[see,][]{agu2009a,agu2009b}.

The stellar and kinematical properties of NLTT~11748 \citep{kaw2009} and of two similar 
objects, LP400-22 
\citep{kil2009,ven2009} and SDSS~J1053+5200 \citep{kil2010}, also suggest that they 
are old halo stars. The constraints placed on their total ages help
retrace the prior evolution of these systems \citep[see][]{tau1999,nel2001,
nel2004}.

The formation of ELM white dwarfs requires that the systems go through
at least two phases of mass transfer, where the second mass transfer phase
strips the less massive companion of its outer envelope before the helium 
ignition. 
This generic scenario appears valid whether 
the more massive component is a neutron star or a white dwarf \citep[see for
example][]{nel2001,nel2004}.
An example of a possible progenitor to an ELM white dwarf is the bright
subluminous B (sdB) star, HD~188112 \citep{heb2003}. The subdwarf star has 
a lower mass than average, $0.24$ versus $\sim0.5\,M_\odot$ \citep{zha2009}, and orbits
a degenerate companion with a mass $\ga 0.73\,M_\odot$ every 14.6 hours.
The subdwarf is a likely progenitor of a helium white dwarf similar to
NLTT~11748.
On the other hand, a possible progenitor to an ELM white dwarf with, this time, a neutron star
companion is SDSSJ102347.6+003841 \citep{wan2009}. This is
a close binary comprising a G-type star and a 1.69ms pulsar
\citep{arc2009} with an orbital period of 4.75 hours. Although the spectral
type of the visible component implies a mass of $\approx1\,M_\odot$, 
the mass ratio of the system dictates
a much lower mass of $\sim 0.2\ M_\odot$ suggesting that the star
is significantly evolved with a helium-enriched core.
This system probably represents a first link between
low-mass X-ray binaries and ELM plus neutron star binaries \citep{arc2009}.

We report new observations that help clarify the nature of the ELM white dwarf 
NLTT~11748.
We obtained two sets of intermediate resolution spectra that show that NLTT~11748 is
in a close binary system (Sect. 2). Our 
analysis of the spectroscopic data and  a first determination of the binary parameters 
 are presented in Sect. 3. We discuss our results in Sect. 4 and
summarise in Sect. 5.

\section{Observations}

We observed NLTT~11748 with the FORS2 
spectrograph attached to the UT1 - Antu at Paranal on UT 2009 October 22 and
23. We used the 1200R grism combined with a slit width of 1'', which provided a 
resolution of 3.0 \AA.
The spectra were obtained as part of a programme aimed at detecting and measuring magnetic
fields in hydrogen-rich white dwarf stars. The observation sequence consists of two consecutive
exposures with the Wollaston prism rotated by 90 degrees between them. 
In principle, a magnetic field may cause
an apparent velocity offset in pairs of exposures, but our analysis
of the radial velocity measurements excludes
the effect of a magnetic field (Sect. 3.1).

We obtained additional observations with the R-C spectrograph
attached to the 4m telescope at the Kitt Peak National Observatory (KPNO) on
UT 2010 March 4 to 6. We used the T2KB detector and
the KPC-24 grating (860 lines/mm) in second order centred on H$\alpha$. 
We employed the GG495 filter to 
block out contamination from the first order. We set the slit width to 1.5'',
which provided a resolution of 0.8\AA\ (full-width half-maximum). 

All spectra were wavelength-calibrated using comparison arc (HeNeAr) 
exposures. The stability of the wavelength scale was verified through
O{\sc i} sky lines, which remained stable with a standard deviation of 3\,km~s$^{-1}$.
All data were reduced with standard procedures within IRAF.
Figure~\ref{fig_obs} and Table~\ref{tbl_log} show our spectra and the observation log.

\begin{table}
\caption{Observation log\label{tbl_log}}
\centering
\begin{tabular}{cccc}
\hline\hline
Start UT date and time & $t_{\rm exp}(s)$ & Start UT date and time & $t_{\rm exp}(s)$ \\
\hline
\multicolumn{4}{c}{ESO} \\
\hline
2009 Oct 22 04:51:00 & 1200 & 2009 Oct 22 06:42:14 & 1200  \\
2009 Oct 22 05:12:09 & 1200 & 2009 Oct 23 04:50:11 & 1200  \\
2009 Oct 22 05:35:52 & 1200 & 2009 Oct 23 05:11:20 & 1200  \\
2009 Oct 22 05:57:00 & 1200 & 2009 Oct 23 05:40:44 & 1200  \\
2009 Oct 22 06:21:05 & 1200 & 2009 Oct 23 06:01:52 & 1200  \\
\hline
\multicolumn{4}{c}{KPNO} \\
\hline
2010 Mar 04 03:19:37 & 1800 & 2010 Mar 05 03:53:24 & 1800  \\
2010 Mar 04 03:50:25 & 1800 & 2010 Mar 05 04:58:15 & 1800  \\
2010 Mar 04 04:25:32 & 1800 & 2010 Mar 06 02:42:39 & 1800  \\
2010 Mar 05 02:53:19 & 1800 &  \\
\hline
\end{tabular}
\end{table}

\section{Analysis}

\subsection{Binary parameters}

\begin{figure}
\includegraphics[width=0.85\columnwidth]{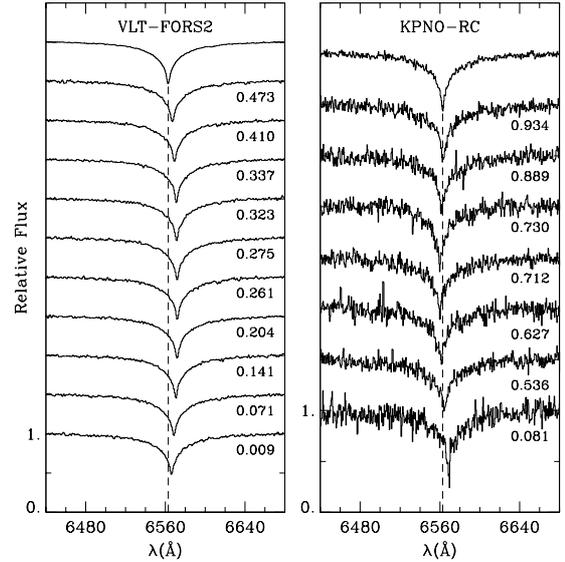}
\caption{Spectra obtained with VLT FORS2 ({\it left})
and the KPNO R-C spectrograph ({\it right}). The spectra are labeled
with the orbital phase (Sect. 3.1) and the top spectra are
the co-added phase-corrected spectrum for each spectrograph.
\label{fig_obs}}
\end{figure}

We used the H$\alpha$ line centre to measure the radial velocity of NLTT~11748.
We fitted a Gaussian profile to the deep, narrow core. Repeated measurements 
varying the fitting window from 2 to 5\AA\ from the line centre were averaged,
leaving residuals of 3 to 10 km s$^{-1}$ depending on the signal-to-noise ratio of the
spectrum. 
The velocities were then corrected to the solar system barycentre. 
We verified the procedure by cross-correlating
each spectrum with a template built by co-adding all spectra in the rest frame (see Fig.~\ref{fig_obs}) 
and comparing the resulting velocities to the Gaussian fits. 
With a window from 6500 to 6620 \AA, we found that the two methods differ by $\Delta \varv=1.4\pm3.5$ km s$^{-1}$
for the higher quality FORS2 spectra, and $\Delta \varv=3.2\pm12.0$ km s$^{-1}$ for the R-C
spectra. Accordingly, the FORS2 and R-C spectra are weighted 4:1 in the orbital analysis.
Table~\ref{tbl_vel} lists the barycentric mid-exposure times and velocities. 

\begin{figure}
\includegraphics[width=0.85\columnwidth]{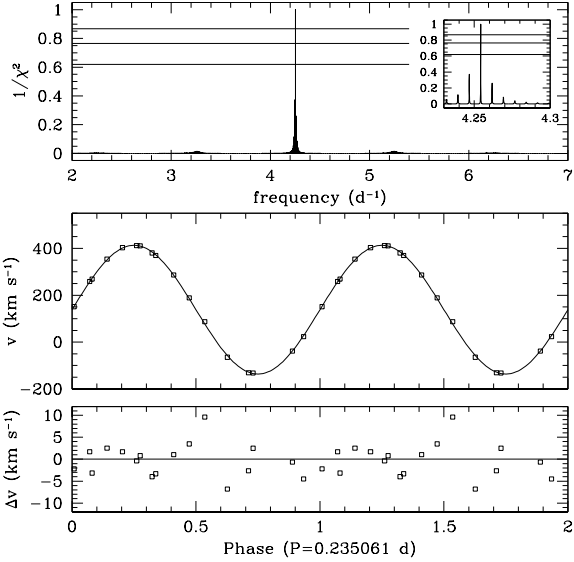}
\caption{({\it Top}) Periodogram of radial velocity measurements showing,
in inset, an enlarged segment near the best period.  ({\it Middle}) Radial 
velocity measurements folded on the orbital period and compared
to a sinusoidal fit (see text). ({\it Bottom}) Residuals of the 
sinusoidal fit.
\label{fig_per}}
\end{figure}

We determined the orbital
parameters by fitting sinusoidal functions to the radial velocities, where
the period (P), mean systemic velocity ($\gamma$), the velocity semi-amplitude
($K$) and the initial epoch ($T_0$) are varied. Figure~\ref{fig_per} shows the
periodogram and best-fit radial velocity curve. The significance values are
drawn at 66\%, 90\% and 99\%. The average of the residuals is $\sim3$ km~s$^{-1}$,
commensurate with our estimates of the wavelength scale accuracy. Although
the FORS2 spectra were obtained with the Wollaston prism in place, the low 
residual value excludes a significant magnetic field.
A more detailed spectropolarimetric analysis will be reported
elsewhere.

\begin{table}
\caption{Radial velocity measurements of NLTT~11748.\label{tbl_vel}}
\centering
\begin{tabular}{cccc}
\hline\hline
BJD  & $v_{\rm bary}$  & BJD  & $v_{\rm bary}$  \\
 (2455000+) & (km s$^{-1}$) & (2455000+) & (km s$^{-1}$) \\
\hline
\multicolumn{4}{c}{VLT/FORS2} \\
\hline
126.71404 & 151.4 &   126.79129 & 369.5  \\
126.72873 & 259.2 &   127.71352 & 412.2  \\
126.74520 & 354.1 &   127.72821 & 380.7  \\
126.75988 & 403.4 &   127.74863 & 286.5  \\
126.77660 & 410.7 &   127.76330 & 189.2  \\
\hline
\multicolumn{4}{c}{KPNO/R-C} \\
\hline
259.64754 &  87.1 &  260.67090 & $-$38.4   \\
259.66893 & $-$64.6 &   260.71593 & 269.0  \\
259.69331 & $-$131.7 &   261.62168 &  23.5  \\
260.62918 & $-$131.1 & &   \\
\hline
\end{tabular}
\end{table}

The best-fit parameters are
\begin{displaymath}
P\ =\ 0.235061\pm0.000003\ {\rm d},
\end{displaymath}
\begin{displaymath}
T_0 ({\rm BJD\ UT})\ = \ 2455126.712\pm0.003,
\end{displaymath}
where $T_0$ is the initial epoch of superior conjunction of the unseen
companion. We also determined the systemic velocity 
\begin{displaymath}
\gamma_{\rm sym} = 136.5\pm0.9\ {\rm km\ s}^{-1}
\end{displaymath}
by subtracting the gravitational redshift of the white dwarf 
($v_g = 2.0\pm0.4$ km~$^{-1}$) from the apparent systemic velocity of
$\gamma = 138.5\pm0.8$ km~s$^{-1}$. The gravitational redshift of the white 
dwarf was calculated assuming the parameters derived in \citet{kaw2009}. The 
observed velocity semi-amplitude is 
\begin{displaymath}
K = 274.8\pm1.5\ {\rm km\ s}^{-1},
\end{displaymath}
which corresponds to a mass function for the unseen companion
\begin{displaymath}
f = 0.5054\pm0.0083\ M_\odot.
\end{displaymath}
Adopting a mass of $0.167\ M_\odot$ for the white dwarf \citep{kaw2009} and using our
orbital parameters, we infer a
minimum mass for the unseen companion of $0.75\ M_\odot$. A main-sequence
companion with a mass of $0.75\ M_\odot$ would have an absolute magnitude of
$M_V \sim 7$ \citep{hen1993}, which would clearly outshine a white dwarf 
with $M_V = 9.7$. 

We re-evaluated the kinematics of the system with the new systemic
velocity to obtain (U,V,W) = (-163,-252,-29) km~s$^{-1}$. These updated values
clearly place NLTT~11748 in the halo. Two other ELM systems whose kinematics 
place them in the halo are LP400-22 \citep{ven2009} and SDSSJ1053$+$5200 
\citep{kil2010} with the remaining systems (without a neutron star)
apparently belonging to the disc \citep{kil2010}.

\subsection{Constraints on a white dwarf companion}

Although \citet{kaw2009} achieved a satisfactory fit to the Balmer line profiles,
their analysis did not consider
a companion. We repeated that analysis using two composite model grids,
DA+DC or DA+DA.
We modelled the primary with a pure-hydrogen model grid (DA) and the secondary with
either a pure-helium (DC) or a DA grid. First, we set the mass of the secondary to $0.75 M_\odot$.
We fitted the Balmer line profiles \citep[see][for details of the NTT-EFOSC2 spectrum]{kaw2009}
varying three parameters: the temperature and surface gravity of the primary ($T_{\rm eff,p},
\log{g}_{\rm p}$) and the effective temperature of the secondary ($T_{\rm eff,s}$). The total flux
was obtained by combining the primary and secondary surface fluxes weighted by the emitting area:
$R^2\,f_{\rm tot} = R_{\rm p}^2 \,f_{\rm p} + R_{\rm s}^2 \,f_{\rm s}$, where $R$ is the
effective radius of the composite star, and $R_{\rm p}$ and $R_{\rm s}$ are the primary
and secondary radii computed using the evolutionary mass radius relations of \citet{ben1999}
and \citet{ser2002}. Note that the flux composition takes into account the calculated orbital phase of the
observed spectrum and the corresponding offset in velocities between the two stars.
The combined spectrum is then fitted to the observed spectrum with
$\chi^2$ minimization techniques. 

The best-fit parameters from the DA$+$DA grid are $T_{\rm eff,p}=8580\pm50$ K ($1\sigma$) and 
$\log{g}_{\rm p}=6.18\pm0.15$ for the primary and $T_{\rm eff,s}\le 9600$ K for the secondary. The 
flux contribution from the secondary at 5500\AA\ (V-band) is $\le 5$\% of the total flux. 
But if we used the DA$+$DC grid we obtained an upper limit of
$T_{\rm eff,s}\le 7200$ K, with a flux contribution $\le 2$\%.
Setting the secondary mass to $1\,M_\odot$ did not
affect our solution for the primary, but allowed a higher secondary temperature ($+400$ K).
In summary, the best-fit parameters are identical to the single
star solution obtained by \citet{kaw2009}, and we conclude that the secondary star is much fainter than the primary star.

\section{Discussion}

If the maximum mass of a white dwarf is $1.35\ M_\odot$,
a white dwarf companion would be expected for inclinations 
$i > 51^\circ$. Ultra-massive white dwarfs have been found to be either
single or in close
binary systems \citep{ven2008}, as in the peculiar binary system RX J0648.0$-$4418 
\citep{mer2009}. 
Constraining the secondary mass between 1.35 and $0.75\,M_\odot$, the system
is not expected to merge before 5.1 to 7.8 Gyr. In the low-mass range the merger 
would create an ultramassive white dwarf, while in the upper mass range the system
would constitute a Type 1a supernova candidate.

\begin{figure}
\includegraphics[width=0.85\columnwidth]{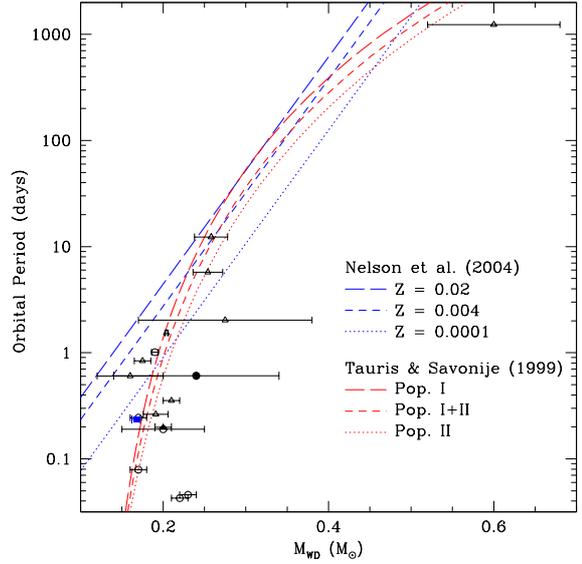}
\caption{Orbital period versus mass of the visible component 
in binary systems containing a white dwarf and a neutron star
({\it open triangles}) or another white dwarf ({\it open circles}, this 
includes systems where the unseen is unknown, but has a high probability of being 
a more massive white dwarf). NLTT~11748, HD~188112, and
SDSSJ1023$+$0038 are marked with a filled square, circle, and triangle, respectively.
These values
are compared to theoretical period-mass relations for NS+WD systems of \citet{nel2004} and 
\citet{tau1999} drawn with thick and thin lines, respectively.
\label{fig_per_mass}}
\end{figure}

Figure~\ref{fig_per_mass} shows the locus of the ELM binary systems in a period versus ELM mass diagram. The data
from Table~\ref{tbl_bin_wd} are compared to the
mass-period relations from the WD+NS binary evolutionary scenarios of
\citet{nel2004} and \citet{tau1999}.
Similar trends 
are also predicted for the brighter star in binary systems with two white 
dwarf stars \citep{nel2001}. 
Figure~\ref{fig_per_mass} also includes the long period WD+NS
system PSR~B0820+02d \citep{koe2000}. This particular system extends the apparent
correlation toward higher mass ($0.60\pm0.08\,M_\odot$) and longer period (1232.5 d).
Table~\ref{tbl_bin_wd} first provides
the parameters of NLTT11748, followed by the parameters of systems 
containing white dwarfs with a neutron star companion (NS+WD), and double white
dwarf systems (WD+WD). Finally, we list the parameters of the possible progenitors
of ELM white dwarf binary systems.
Some scatter between the mass of the brighter star and the orbital period is
clearly visible, but a correlation is still observed between the two 
parameters. The models of \citet{tau1999} appear to better represent the
observed trend than those of \citet{nel2004}. The two outstanding points that
are significantly off the predicted relationship of either models are the two
systems with very short periods 
\citep[SDSSJ1053$+$5200, SDSSJ1436$+$5010:][]{kil2010,mul2010}. Overall, WD+WD and
WD+NS systems overlap in the diagram, so 
this trend alone does not allow one
to distinguish between WD+WD and WD+NS systems.

The ELM white dwarfs that are companions to neutron stars appear to be
very faint ($V > 20$) with the brightest known ELM companion to a neutron star with
an apparent magnitude of $V = 19.6$ (PSR J1012$+$5307). The binary systems
containing an ELM white dwarf and either a confirmed or probable 
white dwarf companion are significantly brighter, with apparent visual 
magnitudes ranging from 16.5 up to 19.3, with NLTT~11748 the brightest of 
the sample. In terms of distances we find that the WD+WD sample lies
within a range of 0.1-1.6 kpc at an average distance of 0.75 kpc, while the WD+NS sample
lies within a range of 0.15-4.4 kpc at an average distance of 1.5 kpc.
Located at a distance of 0.4 kpc and being the brightest known ELM, NLTT~11748  
fits well within
the relatively closer and brighter class of WD+WD systems.
Moreover, NLTT~11748 is not associated with a known radio source, therefore based on these
considerations alone we may conclude that
the companion to NLTT~11748 is another
white dwarf with a mass above average.

Indeed, after this work was submitted for publication, \citet{ste2010} announced the
discovery of primary and secondary eclipses in NLTT11748 confirming the
secondary star as a white dwarf. Their analysis confirms the ELM nature of
the primary star and corroborates our analysis of the binary properties.
Their orbital period measurement, based on eclipse timing, agrees with our spectroscopic
period within error bars. Finally, our ephemeris allows us to identifiy the primary and secondary
eclipses observed by \citet{ste2010} with the corresponding orbital conjunctions.

\begin{table}
\caption{Confirmed binary systems containing a low-mass white dwarf.
\label{tbl_bin_wd}}
\centering
\begin{tabular}{lccc}
\hline\hline
Name & P$_{\rm orb}$ (days) & $M_{\rm WD}$ ($M_\odot$) & Reference \\
\hline
NLTT~11748       & 0.2351  & $0.167\pm0.005$ & 1,2 \\
\hline
\multicolumn{4}{c}{NS + WD} \\
\hline
PSRB1855+09      & 12.3272 & $0.258\pm0.020$  & 3 \\
PSRJ0437-4715    & 5.7410  & $0.254\pm0.018$  & 4 \\
PSRJ0218+4232    & 2.0288  & $0.275\pm0.105$  & 5,6 \\
PSRJ1909-3744    & 1.5334  & $0.204\pm 0.002$ & 7 \\
PSRJ1911-5958A   & 0.8371  & $0.175\pm0.010$  & 8,9 \\
PSRJ1012+5307    & 0.6047  & $0.16\pm0.02$    & 10,11,12 \\
PSRJ1738+0333    & 0.3548  & $0.20\pm 0.05$   & 13 \\
PSRJ0751+1807    & 0.2631  & $0.191\pm0.015$  & 14 \\
\hline
\multicolumn{4}{c}{WD + WD} \\
\hline
LP400-22         & 1.010   & $0.190\pm0.004$  & 15 \\
SDSSJ0822$+$2753 & 0.244   & $0.17\pm 0.01$   & 16 \\
SDSSJ1257$+$5428 & 0.1898  & $0.20\pm 0.05$   & 17 \\
SDSSJ0849$+$0445 & 0.079   & $0.17\pm 0.01$   & 16 \\
SDSSJ1436$+$5010 & 0.0458  & $0.23\pm 0.01$   & 16 \\
SDSSJ1053$+$5200 & 0.0426  & $0.22\pm 0.01$   & 16 \\
\hline
\multicolumn{4}{c}{Progenitors} \\
\hline
HD~188112        & 0.6066  & $0.24\pm0.10$    & 18 \\
SDSSJ1023$+$0038 & 0.1981  & $0.2\pm  0.05$   & 19,20 \\
\hline
\end{tabular}
\tablebib{
(1)~This work; 
(2)~\citet{kaw2009}; 
(3)~\citet{kas1994}; 
(4)~\citet{ver2008}; 
(5)~\citet{nav1995}; 
(6)~\citet{bas2003}; 
(7)~\citet{jac2005};
(8)~\citet{cor2006}; 
(9)~\citet{bas2006};
(10)~\citet{web2004}; 
(11)~\citet{van1996};
(12)~\citet{cal1998}; 
(13)~\citet{fre2008}; 
(14)~\citet{nic2005};
(15)~\citet{ven2009}; 
(16)~\citet{kil2010}; 
(17)~\citet{mar2010};
(18)~\citet{heb2003}; 
(19)~\citet{arc2009}; 
(20)~\citet{wan2009}. 
}
\end{table}

\section{Conclusions}

We find that the ELM white dwarf NLTT~11748 is part of a short-period binary 
system ($P=5.64$ hrs). Based on our orbital analysis alone, we find that the minimum mass of the companion is 
$0.75\,M_\odot$.
However, the companion has independently been identified as a faint white dwarf \citep{ste2010},
and assuming a mass of
$0.75 M_\odot$, we constrained the companion temperature to
$\le 9600$ K (DA) or $\le 7200$ K (DC).
The kinematics of NLTT~11748 also place the system in the Galactic halo.

\vspace{-0.2cm}
\begin{acknowledgements}
A.K. and S.V. are supported by GA AV grant numbers IAA301630901 and 
IAA300030908, respectively, and by GA \v{C}R grant number P209/10/0967. 
We thank the referee for a prompt review and useful suggestions.
\end{acknowledgements}


\begin{thebibliography}{}
\bibitem[Ag{\"u}eros et al.(2009a)]{agu2009a} Ag{\"u}eros, M.~A., Camilo, F.,
Silvestri, N.~M., et al. 2009a, ApJ, 697, 283
\bibitem[Ag{\"u}eros et al.(2009b)]{agu2009b} Ag{\"u}eros, M.~A., 
Heinke, C., Camilo, F., et al. 2009b, ApJ, 700, L123
\bibitem[Archibald et al.(2009)]{arc2009} Archibald, A.M., Stairs, I.H., 
Ransom, S.M., et al. 2009, Science, 324, 1411
\bibitem[Badenes et al.(2009)]{bad2009} Badenes, C., Mullally, 
F., Thompson, S.~E., \& Lupton, R.~H.\ 2009, \apj, 707, 971
\bibitem[Bassa et al.(2006)]{bas2006} Bassa, C.G., van Kerkwijk, M.H., 
Koester, D., \& Verbunt, F. 2006, A\&A, 456, 295
\bibitem[Bassa et al.(2003)]{bas2003} Bassa, C.G., van Kerkwijk, M.H., \&
Kulkarni, S.R. 2003, A\&A, 403, 1067
\bibitem[Benvenuto \& Althaus(1999)]{ben1999} Benvenuto, O.G., \& Althaus, L.G. 1999, \mnras, 303, 30
\bibitem[Callanan et al.(1998)]{cal1998} Callanan, P.J., Garnavich, P.M., \&
Koester, D. 1998, MNRAS, 298, 207
\bibitem[Corongiu et al.(2006)]{cor2006} Corongiu, A., Possenti, A., Lyne, A.G.,
et al. 2006, ApJ, 653, 1417
\bibitem[Freire et al.(2008)]{fre2008} Freire, P.C.C., Jacoby, B.A., \& 
Bailes, M. 2008, AIP Conf. Proc., Vol. 983, 488
\bibitem[Heber et al.(2003)]{heb2003} Heber, U., Edelmann, H., Lisker, T., \&
Napiwotzki, R. 2003, A\&A, 411, L477
\bibitem[Henry \& McCarthy(1993)]{hen1993} Henry, T.J. \& McCarthy, Jr., D.W.
1993, AJ, 106, 773
\bibitem[Jacoby et al.(2005)]{jac2005} Jacoby, B.A., Hotan, A., Bailes, M.,
Ord, S., \& Kulkarni, S.R. 2005, ApJ, 629, L113 
\bibitem[Kaspi et al.(1994)]{kas1994} Kaspi, V.M., Taylor, J.H., \& Ryba, M.F. 
1994, ApJ, 428, 713
\bibitem[Kawka \& Vennes(2009)]{kaw2009} Kawka, A. \& Vennes, S. 2009, A\&A,
506, L25
\bibitem[Kilic et al.(2010)]{kil2010} Kilic, M., Brown, W.R.,
Allende Prieto, C., \& Kenyon, S.J. 2010, ApJ, in press (arXiv:0911.1781)
\bibitem[Kilic et al.(2009)]{kil2009} Kilic, M., Brown, W.~R., 
Allende Prieto, C., et al. 2009, ApJ, 695, L92
\bibitem[Koester \& Reimers(2000)]{koe2000} Koester, D., \& Reimers, D. 2000,
A\&A, 364, L66
\bibitem[Kulkarni \& van Kerkwijk(2010)]{kul2010} Kulkarni, S.~R., \& 
van Kerkwijk, M.~H. 2010, ApJ, submitted (arXiv:1003.2169)
\bibitem[Marsh et al.(2010)]{mar2010} Marsh, T.~R., G\"ansicke, B.~T., 
Steeghs, D., et al. 2010, ApJ, submitted (arXiv:1002.4677)
\bibitem[Mereghetti et al.(2009)]{mer2009} Mereghetti, S., Tiengo, A., 
Esposito, P., et al. 2009, Science, 325, 1222
\bibitem[Mullally et al.(2010)]{mul2010} Mullally, F., Badenes, C., 
Thomson, S.E., \& Lupton, R. 2010, ApJ, 707, L51
\bibitem[Navarro et al.(1995)]{nav1995} Navarro, J., de Bruyn, A.~G., 
Frail, D.~A., Kulkarni, S.~R., \& Lyne, A.~G. 1995, ApJ, 455, L55 
\bibitem[Nelemans et al.(2001)]{nel2001} Nelemans, G., Yungelson, L.~R., 
Portegies Zwart, S.~F., \& Verbunt, F. 2001, A\&A, 365, 491
\bibitem[Nelson et al.(2004)]{nel2004} Nelson, L.~A., Dubeau, E., \& 
MacCannell, K.A. 2004, ApJ, 616, 1124
\bibitem[Nice et al.(2005)]{nic2005} Nice, D.~J., Splaver, E.~M., Stairs, I.~H.,
et al. 2005, ApJ, 634, 1242
\bibitem[Serenelli et al.(2002)]{ser2002} Serenelli, A.M., Althaus, L.G., Rohrmann, R.D., 
\& Benvenuto, O.G.\ 2002, \mnras, 337, 1091 
\bibitem[Steinfadt et al.(2010)]{ste2010} Steinfadt, J.~D.~R., 
Kaplan, D.~L., Shporer, A., Bildsten, L., \& Howell, S.~B.\ 2010, ApJ, in press (arXiv:1005.1977)
\bibitem[Tauris \& Savonije(1999)]{tau1999} Tauris, T.~M. \& Savonije, G.~J. 
1999, A\&A, 350, 928
\bibitem[van Kerkwijk et al.(1996)]{van1996} van Kerkwijk, M.~H., Bergeron, P.,
\& Kulkarni, S.~R. 1996, ApJ, 467, L89
\bibitem[Vennes \& Kawka(2008)]{ven2008} Vennes, S., \& Kawka, A. 2008, MNRAS, 
389, 1367
\bibitem[Vennes et al.(2009)]{ven2009} Vennes, S., Kawka, A., Vaccaro, T.~R., 
\& Silvestri, N.~M. 2009, A\&A, 507, 1613
\bibitem[Verbiest et al.(2008)]{ver2008} Verbiest, J.~P.~W., Bailes, M., 
van Straten, W., et al. 2008, ApJ, 679, 675
\bibitem[Wang et al.(2009)]{wan2009} Wang, Z., Archibald, A.~M., 
Thorstensen, J.~R., et al. 2009, ApJ, 703, 2017
\bibitem[Webb et al.(2004)]{web2004} Webb, N.~A., Olive, J.-F., Barret, D.,
et al. 2004, A\&A, 419, 269
\bibitem[Zhang et al.(2009)]{zha2009} Zhang, X., Chen, X., \& Han, Z.\ 2009, \aap, 504, L13 
\end{thebibliography}
\end{document}